\documentclass[english]{iopart}
\usepackage[T1]{fontenc}
\usepackage[latin9]{inputenc}
\usepackage{graphicx}

\providecommand{\tabularnewline}{\\}
\newcommand{\lyxdot}{.}

\usepackage{iopams}
\usepackage{setstack}

\usepackage{babel}

\begin{document}

\title[Instantaneous time-frequency map signal comparisons]{Comparison of Signals from Gravitational Wave Detectors with Instantaneous
Time-Frequency Maps}

\author{A. Stroeer*, L. Blackburn, J. Camp}
\address{NASA Goddard Space Flight Center, Code 663, 8800 Greenbelt Rd, Greenbelt, MD 20771}
\address{*CRESST, University of Maryland College Park, College Park,
  MD 20742\\now at: University of Texas at Brownsville, Center
  for Gravitational Wave Astronomy, 80 Fort Brown, Brownsville, TX 78520}
\ead{astroeer@phys.utb.edu}
\begin{abstract}
Gravitational wave astronomy relies on the use of multiple detectors,
so that coincident detections may distinguish real signals from instrumental
artifacts, and also so that relative timing of signals can provide
the sky position of sources. We show that the comparison of instantaneous
time-frequency and time-amplitude maps provided by the Hilbert-Huang
Transform (HHT) can be used effectively for relative signal timing
of common signals, to discriminate between the case of identical coincident
signals and random noise coincidences, and to provide a classification
of signals based on their time-frequency trajectories. The comparison
is done with a $\chi^{2}$ goodness-of-fit method which includes contributions
from both the instantaneous amplitude and frequency components of
the HHT to match two signals in the time domain.
This approach naturally allows the analysis of waveforms with strong frequency modulation.
\end{abstract}
\maketitle

\section{Introduction}
\label{sec:Introduction}

\begin{figure}
\includegraphics[scale=0.35]{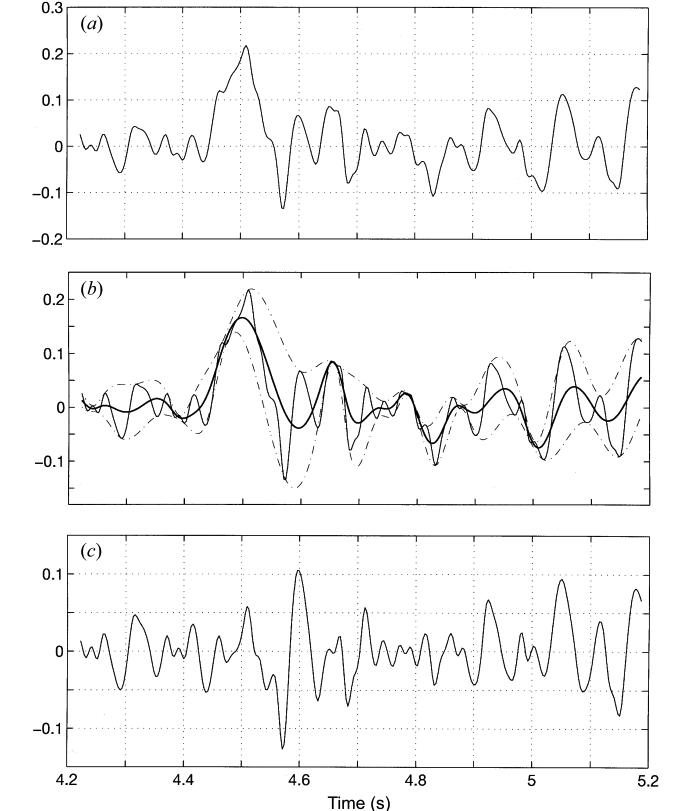}

\caption{Illustration of Empirical Mode Decomposition (EMD). The goal
of the EMD is to decompose a composite time series into the sum of
identifiable oscillatory modes. (a) Original time series. (b) Average of
the upper and lower envelopes formed by fitting extrema. (c) By
subtracting the average from the time series, we extract the highest
frequency mode. This mode is further refined by the same procedure until
a single, stable oscillatory mode is obtained. The mode is removed from
the data (not shown), and the entire procedure is repeated on the
remainder to obtain additional lower frequency modes.}

\label{Flo:emdtheory}
\end{figure}

The Hilbert-Huang Transform (HHT) \cite{huang2005hht} is a novel
data analysis technique used to detect and characterize physical oscillatory
modes in time series data. It is adaptive and data driven, making
it directly applicable to data  containing transient signals and non-stationary noise \cite{huang1998emd}.
It consists of the empirical mode decomposition (EMD), followed by
the Hilbert Spectral Analysis (HSA). The EMD process consists of
forming an envelope about the extrema of the data by cubic spline
fitting, and then subtracting the average of the envelope from the
data. The extrema of remainder are then fitted, and the process is
repeated as many times as necessary to obtain a waveform that is
symmetric about zero mean (within a predetermined tolerance). Once
this has occurred, the waveform is labeled IMF1 and is subtracted from
the original time series, removing the highest frequency content, and
allowing the remainder to be sifted again to obtain the next IMF. This
procedure is illustrated in Figure~\ref{Flo:emdtheory}. The EMD acts
as a dyadic filter, halving the median frequency in each consecutive
IMF \cite{wu2004scw}, while the sum over all the IMFs is the original
data. The narrowband and symmetric form of  the IMFs are essential to
allow the HSA to be applied. 

The HSA derives the instantaneous frequency ($I\! F$) and amplitude
($I\! A$) at each time by differentiating the phase and taking the absolute
value of the analytical complex representation of each IMF, obtained
with the IMF itself (real part) and the Hilbert transform of the IMF
(imaginary part). The IF and IA are then, essentially, the
    frequency 
and amplitude of the best fit sinusoidal wave at each point in time \cite{Cohen:1995:TAT:200604}. Through this procedure the HHT allows extraction
of instantaneous time-frequency and time-amplitude trajectories, in
contrast to integral methods like Fourier or Wavelet analysis which
are limited fundamentally by time-frequency spreading \cite{HuangIFAADA}.
This instantaneous character of the HHT maps allows for high resolution
time-frequency analysis, and in particular allows the detailed, high
precision study of waveforms with strong frequency modulation. The
HHT has been applied to the time series analysis of heart rate monitors
\cite{EKG}, studies of electroencephalograms \cite{1741-2552-7-4-046008},
investigations into the integrity of structures \cite{HHTexample2},
and measurements of the seismic surface wave dispersion \cite{TAOhhtexample}. 

Time series analysis is now used in the search for gravitational wave
(GW) sources. A number of high sensitivity detectors, including the
US Laser Interferometer Gravitational Wave Observatory (LIGO)\cite{abramovici1992lli,barish1999lad},
consisting of a 4 km detector in Louisiana and a 4 km and a 2 km detector
in Washington, the European Virgo Observatory \cite{bradaschia1990vpw},
consisting of a 3 km detector in Pisa, Italy, and the GEO600
observatory \cite{geodetector} consisting of a 600m detector near
Sarstedt, Germany, are now in operation
and analyzing data. A key concept in the data analysis is that the
coincident appearance of signals in two or more detectors will be
important in discriminating a real GW signal from an instrumental
artifact. Signal comparison is also required in order to establish
the relative timing of the appearance of an event in multiple widely-separated
detectors so that the sky position of a possible source may be determined.

In this paper we demonstrate the use of instantaneous time-frequency
maps provided by the HHT for relative signal timing, for use as a
veto of accidental noise coincidences, and to characterize the time-frequency
structure of transient instrumental noise artifacts (known as {}``glitches'')
in recorded gravitational wave data. In all these studies frequency
modulation plays an important role. We show these results using gravitational-wave
data from the fourth science run of the Laser Interferometer Gravitational
wave Observatory (LIGO S4) \cite{0264-9381-25-24-245008}, which took
place in early 2005. The S4 data run spanned two months and had detector
sensitivities within a factor of two of the design goal. 

The paper is organized as follows: in Sec. \ref{sec:The-Hilbert-Huang-Transform}
we summarize the application of the HHT to this study. We show empirically
in Sec. \ref{sec:A--approach} that the measurement errors in the
instantaneous amplitude and frequency provided by the HHT are a simple
function of the local signal-to-noise ratio and are approximately
Gaussian. This allows the use of a $\chi^{2}$ goodness-of-fit method
to evaluate the match between time-frequency and time-amplitude maps,
which can be applied to simulated signals as well as noise
transients in S4 data. Sec. \ref{sec:Timing-with-the} shows the performance
of relative timing between two common signals in noise obtained by
minimizing the $\chi^{2}$ as a function of relative time delay. The
use of the $\chi^{2}$ test to veto similar but unequal waveforms that could
be caused by coincident noise transients
is studied in Sec. \ref{sec:Vetoing-unequal-waveforms}, and grouping
of instrumental artifacts into classes based on time-frequency and
time-amplitude structure (glitch morphology) is discussed in Sec.
\ref{sec:A-glitch-morphology}.

\section{Application of the Hilbert-Huang Transform in this work\label{sec:The-Hilbert-Huang-Transform}}

We use an application of the HHT algorithm as described in \cite{2009PhRvD..79l4022S}
to analyze GW data with low integrated signal-to-noise ratio, SNR
= $\sqrt{\sum_{i}h_{i}^{2}}/s$, where 
$s$ is the standard deviation of the discrete stationary noise
in the time series after whitening and $h_{i}$ is the discrete (whitened) measured strain
of the signal. This application uses Bayesian
blocking \cite{mcnabb2004obe} to identify regions of excess power
in the instantaneous amplitude ($I\! A$), and uses the FFT-derived power
spectrum of these regions to establish the maximum frequency of the
event. It then low-pass filters the data at a frequency close to this
maximum frequency to reduce distortion of the data from noise at frequencies
greater than those contained in the signal. It performs a final processing
technique (EEMD, \cite{wu2005eem}) to average over errors in the
EMD process.

In this study we have made the following modifications to the HHT
application to the GW data described above: 
\begin{itemize}
\item The extrema of
a time-sampled signal can occur between two distinct sampling points
and thus may not be present in the discretized time series.
The standard EMD uses the sampling point closest to the true extrema as
the contact point for the upper envelope, leading to a false offset
in magnitude and time. 
In this case, the mean of the data is underestimated, which leads
to a decomposition error in the final IMF. This error manifests as
a frequency modulation in the $I\! F$ and an amplitude modulation in the
$I\! A$, seen as a sinusoidal oscillation around the true value with a
period equal to the separation of the extrema in time. To correct
for this we interpolate the data around the extrema with a cubic spline,
and derive analytically the zero crossing of the first derivative
of the spline in order to establish the true extrema of the data.
The interpolated value in time and magnitude replaces the false extrema.
\item For more stable results, and to capture the complete frequency content
 of an event within the dyadic frequency range of the first IMF, we
have found that it is best to use a low-pass filter with frequency
cut-off at twice the estimated maximum Fourier frequency of the signal. 
\end{itemize}

\section{Comparison of signals: instantaneous time-amplitude and time-frequency
maps, associated errors, and their use in a $\chi^{2}$ goodness-of-fit
measure\label{sec:A--approach}}

A quantitative comparison of time-frequency maps of signals from different
detectors can form a basis for estimating relative signal timing,
and the likelihood of the signals being consistent. We approach the
problem of comparison of signals from different detectors through
the use of a $\chi^{2}$ goodness-of-fit measure between the instantaneous
time-amplitude and time-frequency maps of the signals generated by
the application of the HHT to the time series data. 

\begin{figure}
\includegraphics[angle=-90,scale=0.55]{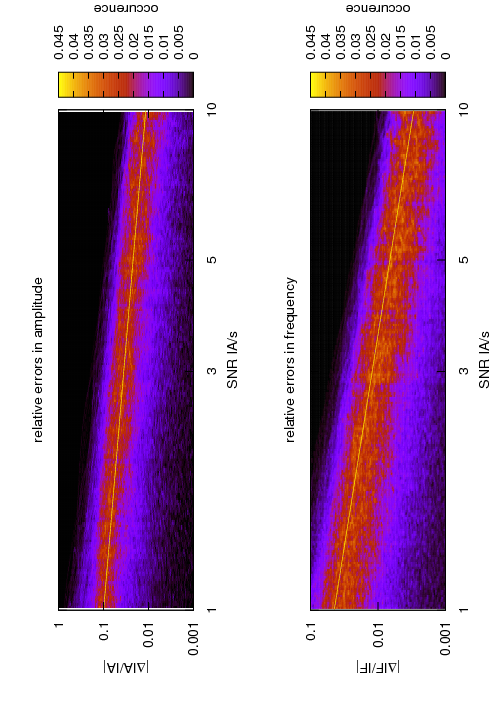}

\caption{2-D histogram of the absolute fractional uncertainties in the
extraction of instantaneous frequencies and amplitudes of the HHT as a
function of local signal-to-noise (SNR = $I\!A/s$) for a population of
sine-Gaussian, BBH merger, and white noise burst signals. Occurence
represents the percent occupancy of the population at fixed SNR. The
1$\sigma$ spread in the fractional measurements correspond to the most
likely absolute fractional error in the log-scale binning on $|\Delta
I\!F/I\!F|$. We find that the 1$\sigma$ uncertainties in $I\!A$ and $IF$ are
well described by a power-law in normalized amplitude $I\!A/s$.}

\label{Flo:TIMING}
\end{figure}
 
In this approach the degree of match of two signals from detectors
1 and 2 is determined by computing a reduced $\chi^{2}$
statistic using $I\! A$ and $I\! F$ ($N$ denoting the total number of sampling points
in time) \begin{equation}
\chi_{1,2}^{2}=\frac{1}{2N}\left[\sum_{n=1}^{N}\frac{(I\! A_{1}[n]-I\! A_{2}[n])^{2}}{\sigma_{I\! A_{1}}^{2}[n]+\sigma_{I\! A_{2}}^{2}[n]}+\sum_{n=1}^{N}\frac{(I\! F_{1}[n]-I\! F_{2}[n])^{2}}{\sigma_{I\! F_{1}}^{2}[n]+\sigma_{I\! F_{2}}^{2}[n]}\right],\end{equation}
where $\sigma[n]$ is the uncertainty in the $I\! A$ or $I\! F$, both of which
are time-dependent functions of $I\! A$. For this test, $I\! A$ is also
normalized by its maximum absolute value so that the test is sensitive only
to the shape of the $I\! A$ evolution and not the absolute magnitude. This is
important, for example, in the case of a gravitational-wave
which may be subject to different antenna scaling factors from misaligned detectors, as
well as amplitude errors in detector calibration.
$2N$ does not denote the effective
number of degrees of freedom in this reduced $\chi^2$ because of
correlations between the data points, which are generally sampled
at a much higher rate than twice the bandwidth of data contained in the
IMFs. In practice an empirical threshold on the reduced $\chi^2$ is
determined through Monte Carlo studies to limit the false-dismissal
rate for identical signals to 1$\%$ or less.

The fractional uncertainty for the $I\! A$ and $I\! F$ produced by the HHT is
related at each point in time to the measured ratio $I\! A/s$, where $s$
here refers to the standard devation of the white noise within the IMF band. In
general a larger $I\! A/s$ yields a smaller fractional uncertainty.
We determine these uncertainties by injecting
a variety of ad-hoc signals used in the LIGO S4 burst search \cite{0264-9381-25-24-245008}
as well as waveforms from numerical relativity into white Gaussian noise, with the SNR of
these signals ranging from 1 to 30. These signals include sine-Gaussians
with frequencies ranging from 60 Hz to 3 kHz
and quality factors, Q (number of cycles), from 3 to 9, 
two equal mass binary black hole (BBH) merger signals \cite{baker2006bbh,baker2006gwe}
with total system masses of 20 and 60 $M_{\odot}$ and ring-down
frequencies (oscillation frequency of the resulting black hole after merger) of 400 and 250
Hz respectively, and white noise bursts consisting of band passed noise with a flat spectral density between 100 and
200 Hz and a Gaussian time-envelope with a width of 1, 10 and 100 ms. We then use
the HHT to extract the $I\! A$ and $I\! F$ for each signal, and obtain the
fractional uncertainty by comparing the $I\! A$ and $I\! F$ with the values
seen in zero noise (see Fig. \ref{Flo:TIMING}). The $1\sigma$ $I\! A$ uncertainty
and the $1\sigma$ $I\! F$ uncertainty versus measured $I\! A/s$ is found by fitting a straight line to the
most likely absolute fractional uncertainty in the log-log plane, so that the fractional uncertainty takes
the form of a power law in $I\! A/s$,
\begin{equation}
\log_{10}(\sigma_{I\! A})\approx-1.12\cdot\log_{10}(I\! A/s)-0.52\label{eq:errorIF}\end{equation}
\begin{equation}
\log_{10}(\sigma_{I\! F})\approx-0.60\cdot\log_{10}(I\! A/s)-0.28\end{equation}

\begin{figure}
\includegraphics[angle=-90,scale=0.5]{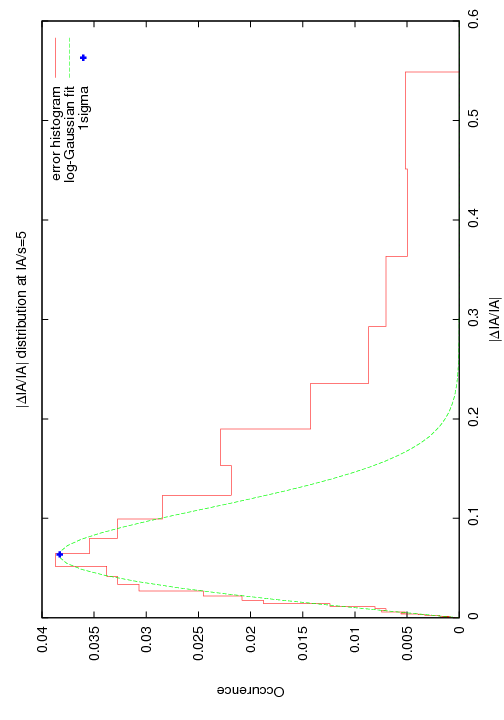}

\caption{Histogram of absolute fractional uncertainties in IA for a
  population of simulated signals at IA/s=5, and a one-sided Gaussian
  fit to the distribution with corresponding 1$\sigma$ indicated. 
Due to the log-scale binning in $|\Delta IA/IA|$, the 1$\sigma$ spread
in the one-sided distribution 
corresponds to maximum occurance. The histogram is normalized to 1.}

\label{Flo:1Derror}
\end{figure}

We assign these uncertainties to the extraction of $I\! A$ and $I\! F$ from
LIGO data by referring the ratio $I\! A/s$ at each data point to
these linear fits. We find the uncertainties to be roughly Gaussian
distributed at any value of $I\! A/s$, but to also have non-Gaussian
features towards higher $I\! A/s$ (see Fig. \ref{Flo:1Derror}).

\section{Timing with the $\chi^{2}$ statistic\label{sec:Timing-with-the}}

The relative timing of signals from a network of gravitational wave
detectors can be used to determine the sky location of the astrophysical
source of a gravitational wave signal (e.g. \cite{PhysRevD.78.122003})].
The $\chi^{2}$ goodness-of-fit measure can be used for timing the
arrival of common signals by time shifting one detector time series
with respect to the other by small intervals up to a total of $\pm10$
ms (the maximum travel time of a GW between the LIGO sites), calculating
the $\chi^{2}$ for each time-shift, and noting the relative time-shift
corresponding to the minimal $\chi^{2}$ value. This minimum then
denotes the best overlap of the time-frequency and time-amplitude
maps of the two detector data streams.

\begin{figure}
\includegraphics[scale=0.38]{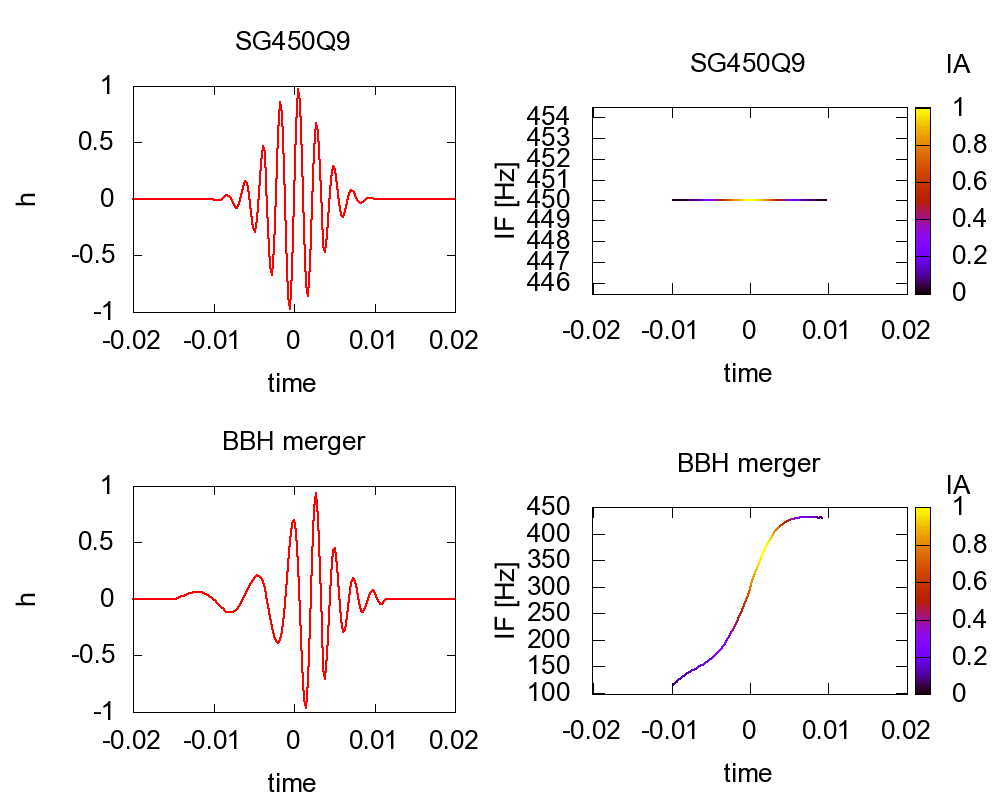}

\caption{Signal waveforms and HHT-derived time-frequency maps for the signals
used to test relative timing: sine-Gaussian f=450, Q=9 and a 20 $M_{\odot}$
BBH merger. The 66.7, 95.7, and 99.7 percentiles of the distribution of absolute
time-shift error versus SNR are shown in the top panel.}

\label{Flo:wave}
\end{figure}
 
\begin{figure}
\includegraphics[scale=0.50]{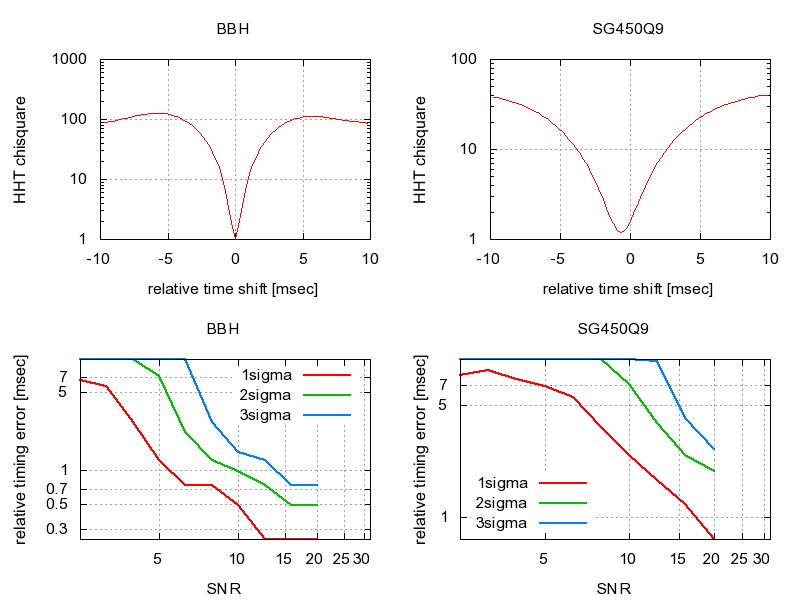}

\caption{Top panels: $\chi^{2}$ comparison of time-frequency and time-amplitude
maps versus time-shift for a single SNR=20 $M_{\odot}$ BBH merger (left)
and sine-Gaussian 450Hz Q9 (right) waveform injected in both
detectors. 
The true difference in arrival time is estimated by locating the
minimum $\chi^2$ value. Lower panels: the aggregate relative timing errors
obtained for these waveforms as a function of SNR. The 
66.7, 95.7, and 99.8 percentiles for the distribution of absolute timing errors
are shown for an ensemble of injections.}

\label{Flo:errortimelag}
\end{figure}

To demonstrate this technique we use two test waveforms: the binary
black hole merger with total mass 20 $M_{\odot}$  and a sine-Gaussian
of frequency 450 Hz and Q of 9. The waveforms and instantaneous time-frequency
maps of these signals are shown in Fig. \ref{Flo:wave}.
Fig. \ref{Flo:errortimelag} shows the reduced
$\chi^{2}$ versus time-shift for the two waveforms when injected
into different realizations of Gaussian white noise at SNR = 20. The
injections were also performed multiple times with varying SNR.
At each SNR, we note the spread
of the relative timing measurements obtained by minimizing the $\chi^2$ versus time-shift. The 66.7, 95.7, and 99.7 percentiles
(1, 2, and 3$\sigma$ respectively) of the distribution of absolute
time-shift error versus SNR are shown in Fig. \ref{Flo:errortimelag}.
As expected, we observe an improvement in timing with rising SNR.
Overall the binary black hole merger shows superior relative timing to the sine-Gaussian
waveform as the sine-Gaussian's flat frequency profile
at 450 Hz reduces the contrast of the $\chi^{2}$ with respect to
small time-shifts. For the back hole merger waveform at SNR=10, 
the 1$\sigma$ relative timing error between two detectors is $\sim$0.5 
ms, 
which is comparable with matched filter methods
\cite{2010arXiv1004.4537V}.

\section{Vetoing unequal waveforms with instantaneous time-frequency maps\label{sec:Vetoing-unequal-waveforms}}

\begin{figure}
\includegraphics[scale=0.45]{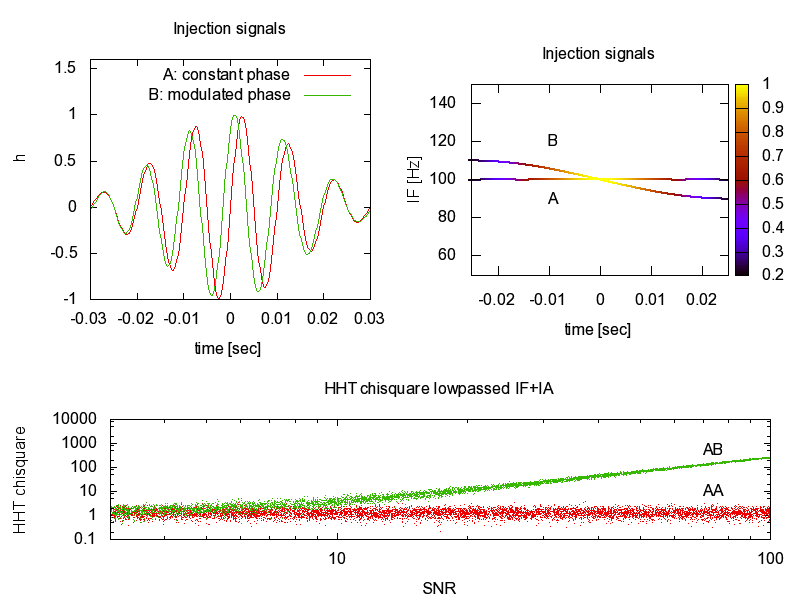}

\caption{The HHT $\chi^{2}$ goodness of fit is used to differentiate very
similar waveforms, for use in veto strategies of instrumental noise
coincidences between two gravitational wave detectors. We display
the $\chi^{2}$ comparison of two waveforms, with one of constant
frequency (A) and the other phase modulated at 1/10th of the base
frequency (B). The two waveforms have a linear correlation coefficient
of 0.96. The top left panel shows the waveforms. The top right panel
shows the $I\! F$ of each waveform versus time, with $I\! A$ color coded. The
lower panel is a scatter plot of reduced $\chi^{2}$ values between
waveforms A and B (green) and identical waveform A and A (red) as
a function of SNR. By applying a reduced $\chi^{2}$ threshold of
2.7, waveforms A and B can be distinguished 63\% of the time at SNR=7,
and 93\% of the time at SNR=10 while maintaining a false-dismissal
probability of 1\% or lower for identical waveforms. }

\label{Flo:veto}
\end{figure}
 
The HHT $\chi^{2}$ goodness of fit applied to $I\! A$ and $I\! F$ can also be used to veto
similar but unequal waveforms. Thus it may be used for veto strategies
that seek to reject instrumental noise coincidences of signals which
are closely related but not identical. To illustrate the HHT $\chi^{2}$
veto, we consider the following very similar waveforms:
\begin{itemize}
\item waveform A: $\sin\left(2\pi100t\right)\exp\left[\frac{(2\pi100t)^{2}}{2\cdot9^{2}}\right]$
\item waveform B: $\sin\left(2\pi100t+\cos(2\pi10t)\right)\exp\left[\frac{(2\pi100t)^{2}}{2\cdot9^{2}}\right]$
\end{itemize}
The second waveform is chosen to have a 10 Hz cosine phase modulation
with respect to the 100 Hz base frequency of the first, as low
frequency modulations about a 100 Hz base frequency are often seen in
instrumental glitches (see Section 6). Both waveforms have
a Gaussian amplitude and a Q$\sim$9. The linear correlation coefficient
$\left[\sum{(h_A-\bar{h}_A)(h_B-\bar{h}_B)}\right]/\sqrt{\sum{(h_A-\bar{h}_A)^2}\sum{(h_B-\bar{h}_B)^2}}$
of the waveforms is 0.96 (0 is uncorrelated while 1 is fully correlated),
indicating a very high degree of similarity. These waveforms are
injected into white Gaussian noise at varying SNR (see Fig. \ref{Flo:veto}).
By setting a threshold in the reduced $\chi^{2}$ of 2.7, the waveforms
A and B can be shown to be dissimilar and thus vetoed 63\% of the
time at SNR = 7, vetoed 93\% of the time at SNR~=~10, and vetoed essentially
100\% of the time at SNR~=~15. The false rejection probability, determined
by applying this veto procedure to identical waveforms (i.e. A and
A) is 1\% or less over the full range of SNR. 

\section{Glitch morphology of S4 data\label{sec:A-glitch-morphology}}

In this section we use the comparison of time-frequency maps of instrumental
noise artifacts (known as {}``glitches'') from the GW channel of
the Livingston 4 km detector (L1) during S4 in order to group them into classes whose
members contain similar time-frequency structure; the goal is that
these groupings may correspond to specific physical mechanisms which
cause the disturbance. We refer to this grouping as establishing a
{}``morphology'' of the glitches. Instrumental noise artifacts
are first identified in the S4 science data (after data quality \cite{0264-9381-25-24-245008} cuts have been applied)
by using the KleineWelle (KW) algorithm \cite{Chatterji:2004qg} which
searches for excess signal power in the wavelet domain.
Triggers are required to be isolated with no other trigger within 1 second to
simplify signal comparison, and are also required to be 
of sufficient strength (SNR$\gtrsim$8.7) for precise time-frequency reconstruction.
We apply the $\chi^{2}$
test on these events in order to find glitches with similar time-frequency
and time-amplitude trajectories, which we then group into the same
morphology class. We consider two events to be in the same class if
their reduced $\chi^{2}$ is less than 2. 

\begin{figure}
\includegraphics[scale=0.3]{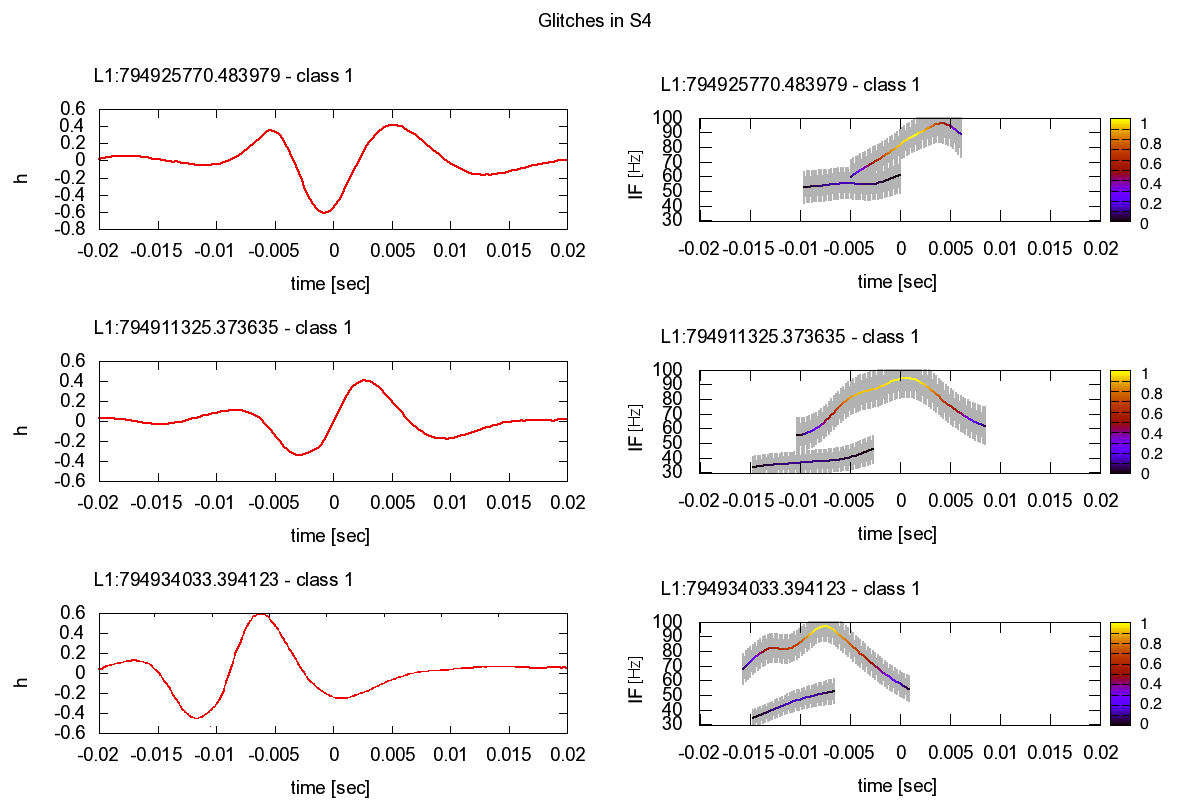}
\caption{Three members of glitch morphology class 1 are displayed with
  their time-frequency
maps. $I\! A(t)$ is indicated by the color scale, and uncertainties
in measured $I\! F$ are indicated as grey shading. The time-frequency structure of the two
modes of this glitch class is present throughout the 31 members.}
\label{Flo:morphFIN}
\end{figure}

To illustrate this approach, we present four of the 14 most populated glitch classes, summarized in Table \ref{Flo:tab}. Each class contains more than 10 individual members, and are chosen for their
degree of match across
class members or common association with auxiliary interferometric channels. For example we show in Fig. \ref{Flo:morphFIN} three members of class 1. The similar time-frequency structure is present throughout the 31 members of this class.
As seen in these three examples, the glitches can show complicated
structure consisting of more than one oscillatory mode. These modes are
separated by the EMD procedure, and they are shown by separate
traces on the time-frequency maps.
The weaker mode in this particular example is centered
at 50 Hz with fairly low
modulation, while the stronger mode shows stronger frequency
modulation peaking at around 100 Hz. The detailed frequency modulation of the modes is represented
by the time-frequency traces, with errors (obtained
through the application of Eq. \ref{eq:errorIF}) shown as grey
bands. 

\begin{figure}
\includegraphics[scale=0.3]{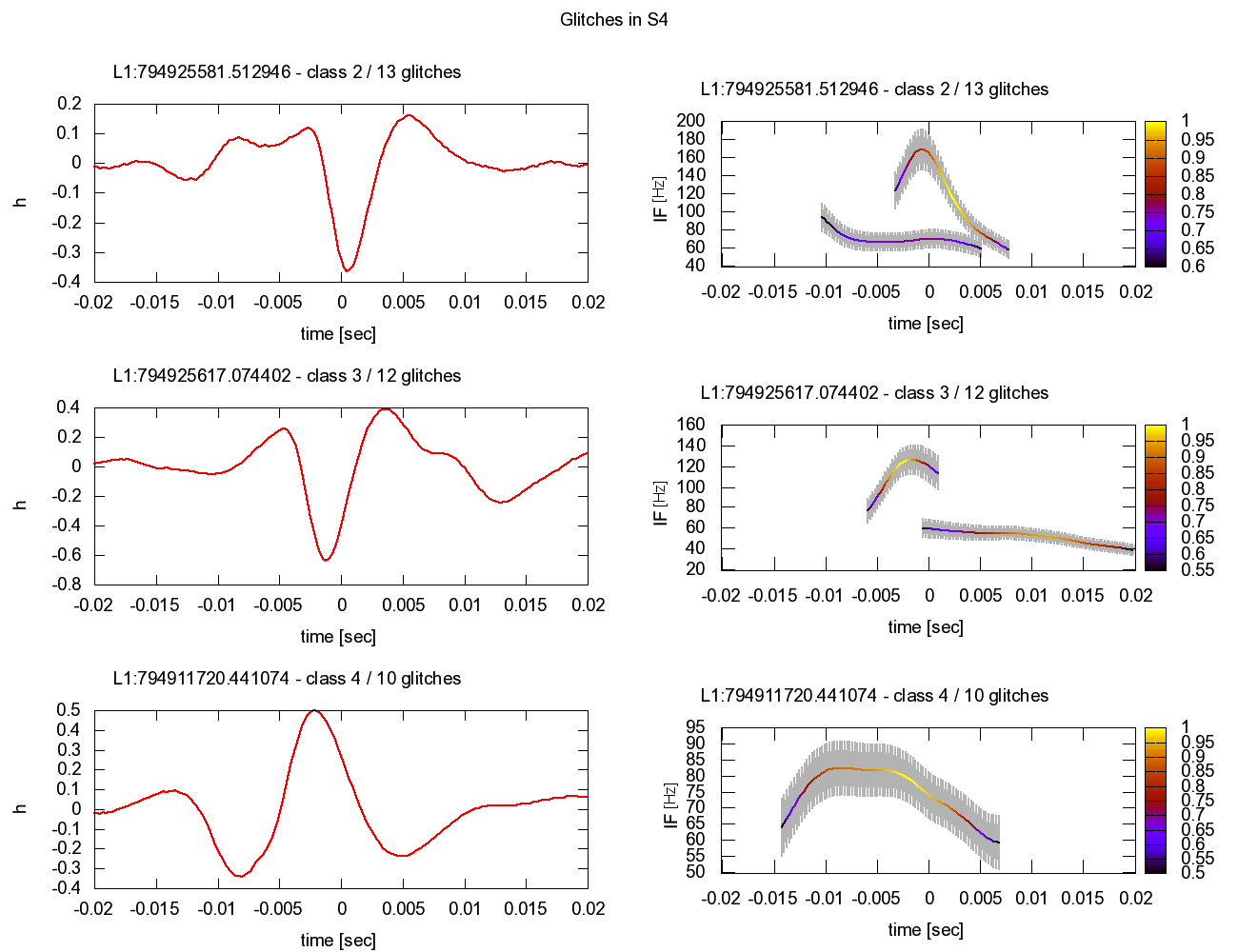}

\caption{An individual example from each of three different glitch classes which show evidence of a link to auxiliary channels.
Left panels show the whitened time series, right panels show the time-frequency
maps with $I\! F(t)$ with $I\! A(t)$ indicated by the color scale, and uncertainties
in measured $I\! F$ indicated as grey shading. }

\label{Flo:morphLINK}
\end{figure}

We also show in Fig. \ref{Flo:morphLINK} a comparison of the time-frequency structures of the glitches from three additional classes 2, 3, and 4.
The time-frequency
map of class 2 in Fig. \ref{Flo:morphLINK} consists of two
physical modes, one around 60 Hz with a shallow frequency modulation,
and one peaking at 180 Hz with a stronger modulation. These modes
can also be seen in the time series of the glitch, where a slower
oscillation is superimposed on top of a faster, larger oscillation
around $t=0$. Class 3 consists of two physical modes
at 120 and 60 Hz, while class 4 appears to be a single physical mode
at 80 Hz.

After establishing glitch morphology classes, we look for correlations of the different classes
with auxiliary channels.
This is done by looking for time coincidences
of KW auxiliary channel triggers within a 50 ms window about
the peak time of the GW channel trigger.
In morphology classes 2, 3 and 4, more than 50\% of the members
can be associated with a simultaneous transient disturbance in a channel related
to the control of the interferometer (PRC), a channel
related to the alignment of an interferometer test mass (WFS), or
a channel related to the intensity of light picked off from the beam
splitter (POB). These correlations are detailed in
Tab. \ref{Flo:tab}. 

The overall goal of grouping of the glitches into these classes is to
obtain insight into the physical processes producing them. 
So far achieving this goal remains elusive in S4 data. The modes and frequency
modulations shown in Fig. \ref{Flo:morphFIN} do not easily suggest any
physical mechanism. Nevertheless, their common time-frequency
structure indicates some underlying cause. We believe that the cataloging of
groups of glitches based on time-frequency structure, and the ongoing
correlation of these classes with auxiliary channels, is a useful first step
to understanding and ultimately removing them.

\begin{table}
\begin{tabular}{|c|c|c|}
\hline 
Morphology class & member count & remarks\tabularnewline
\hline
\hline 
1 & 31 & 60 \% PRC; 52 \% POB \tabularnewline
\hline 
2 & 13 & 62 \% PRC; 54 \% POB\tabularnewline
\hline 
3 & 12 & 50\% WFS, 50\% POB\tabularnewline
\hline 
4 & 10 & 50 \% POB\tabularnewline
\hline
\end{tabular}
\caption{A glitch morphology study on S4 data in the Livingston 4km
  detector. Four of the 14 most populated glitch morphology classes are tabulated
which demonstrate consistent time-frequency structure or association
with auxiliary interferometric channels. Individual members from each
class can be found in Figures \ref{Flo:morphFIN} and \ref{Flo:morphLINK}.}

\label{Flo:tab}
\end{table}

\section{Conclusions\label{sec:Conclusions}}

In this study we showed how instantaneous time-frequency and time-amplitude
maps, as provided by the Hilbert-Huang Transform, can be used to determine
the relative timing of signals in a network of gravitational-wave
detectors, to enable
a veto that can discriminate between similar but unequal waveforms,
and to group instrumental artifacts based on their time-frequency
trajectories. The $\chi^{2}$ goodness-of-fit test proposed provides a measure
of comparison of the time-frequency and time-amplitude maps. The high
time resolution of the instantaneous maps allows strong frequency
modulations to be addressed in a straightforward way, a critical factor
in gaining insight into the physical nature of the signals and instrumental
artifacts. 

Future research will concentrate on the errors in the HHT $I\! F$ and $I\! A$
extraction, with the goal of understanding their origin and minimizing
their magnitude. These errors impact the effectiveness of the timing
and veto estimates directly, and also affect the level of discrimination
of the glitch morphology groupings. 

\section*{Acknowledgements}
The authors would like to thank Gabriela Gonzalez and
Michele Zanolin for
helpful comments and suggestions relating to this work.

\section*{References}

\bibliographystyle{unsrt.bst}
\bibliography{hhtatlowsnr}

\end{document}